\documentclass{jfm}
\usepackage{graphicx}
\usepackage{epstopdf, epsfig}
\usepackage{amsmath,color,graphicx}
\usepackage{ar}

\shorttitle{Ground Effect Scaling}
\shortauthor{A. Mivehchi et al.}

\title{Scaling Laws for the Propulsive Performance of a Purely Pitching Foil in Ground Effect}

\author{Amin Mivehchi\aff{1}
  \corresp{\email{mivehchi@lehigh.edu}},
  Qiang Zhang\aff{2}, Melike Kurt\aff{1}, Daniel B. Quinn\aff{2}
 \and Keith W. Moored\aff{1}}

\affiliation{\aff{1}Department of Mechanical Engineering and Mechanics, Lehigh University,
Bethlehem, PA 18015, USA
\aff{2}Department of Mechanical Engineering, University of
Virginia, Charlottesville, VA 22094, USA}

\definecolor{cadmiumgreen}{rgb}{0.0, 0.42, 0.24}

\begin{document}

\maketitle

\begin{abstract}
Scaling laws for the thrust production and power consumption of a purely pitching hydrofoil in ground effect are presented.  For the first time, ground effect scaling laws based on physical insights capture the propulsive performance over a wide range of biologically-relevant Strouhal numbers, dimensionless amplitudes, and dimensionless ground distances. This is achieved by advancing previous scaling laws \cite[]{moored2018inviscid} with physics-driven modifications to the added mass and circulatory forces to account for ground distance variations.  The key physics introduced are the increase in the added mass of a foil near the ground and the reduction in the influence of a wake vortex system due to the influence of its image system.  The scaling laws are found to be in good agreement with new inviscid simulations and viscous experiments, and can be used to accelerate the design of
bio-inspired hydrofoils that oscillate near a ground plane or two out-of-phase foils in a side-by-side arrangement.

\end{abstract}

\section{Introduction}
In nature, animals such as birds and flying fish use steady ground effect to improve their cost of transport or gliding distance \citep{hainsworth1988induced, rayner1991aerodynamics,park2010aerodynamic}. Similarly, some fish exploit \textit{unsteady} ground effect to improve their cost of transport or cruising speed when swimming near substrates and sidewalls \citep{blake1983mechanics, webb1993effect, webb2002kinematics, nowroozi2009whole, blevins2013swimming}. In unsteady ground effect, fin/wing/tail/body oscillations create time-dependent wakes and time-varying fluctuations in the pressure field that are altered from those of an isolated swimmer or flyer. 

Unsteady ground effect was first examined through the development of analytical models for a fluttering plate in a channel \cite[]{tanida2001ground} and for an oscillating wing in weak ground effect \cite[]{iosilevskii2008asymptotic}, but these only apply in extreme cases, such as flying/swimming very far from or very close to the ground.  At more moderate ground proximities, experiments and computations have shown that rigid \citep{ quinn2014unsteady, Mivehchi2016, perkins2017rolling} and flexible \citep{blevins2013swimming, quinn2014flexible, fernandez2015large, dai2016self, park2017hydrodynamics, zhang2017free} oscillating foils and wings can have improved thrust production with little or no penalty in efficiency when operating in unsteady ground effect. Additionally, \citet{kurt2019swimming} reported the presence of a stable equilibrium altitude for a freely-swimming pitching foil in the presence of a ground plane that was previously observed in the lift force of constrained flapping foils both experimentally \citep{ Mivehchi2016, perkins2017rolling}, and numerically \cite[]{quinn2014unsteady,kim2017autonomous}.

To understand the origins of thrust and efficiency in unsteady ground effect, we can rely on scaling laws. The basis of many recent scaling laws lies in classic unsteady linear theory. The theories of \cite{theodorsen1935general}, \cite{garrick1936propulsion}, and \cite{von1938airfoil} have become particularly useful in this pursuit due to their clear assumptions (incompressible and inviscid flow, small-amplitude motions, non-deforming, and planar wakes) and the identification of the physical origins of their terms.  For instance, these theories decompose the forces acting on unsteady foils into three types: added mass, quasi-steady, and wake-induced forces. Theodorsen's theory was extended by \cite{garrick1936propulsion} by accounting for the singularity in the vorticity distribution at the leading edge to determine the thrust force produced and the power required by such motions. By following \cite{garrick1936propulsion}, \cite{dewey2013scaling} and \cite{quinn2014unsteady,quinn2014scaling} scaled the thrust forces of pitching and heaving flexible panels with their added mass forces. \cite{moored2018inviscid} advanced this previous work by considering the circulatory and added mass forces of self-propelled pitching foils as well as wake-induced nonlinearities that are not accounted for in classical linear theory \cite[]{garrick1936propulsion}. It was shown that data generated from a potential flow solver and from experimental measurements \cite[]{ayancik2019scaling} were in excellent agreement with the proposed scaling laws. Similarly, \cite{floryan2017scaling} considered both the circulatory and added mass forces and showed excellent collapse of experimental data with their scaling laws for the thrust and power of a heaving or pitching two-dimensional rigid foil.  While these studies have provided great insights into the origins of unsteady force production, they were limited to isolated propulsors.

Here, we advanced the scaling laws for isolated purely pitching propulsors developed by \cite{moored2018inviscid} to account for the proximity to the ground.  These scaling laws provide new insight into the underlying physics of unsteady ground effect and are verified through simulations and experiments. Furthermore, we show that the added mass forces of the core two-dimensional scaling relations \cite[]{moored2018inviscid} can be modified by accounting for the increase in the added mass of an object near a ground plane derived from classical hydrodynamic theory \cite[]{brennen1982review}, and the circulatory forces can be modified by accounting for bound and wake vortex-body interactions in ground effect. The newly developed scaling laws offer a physical rationale for the origins of force production, power consumption, and efficient unsteady swimming in proximity to the ground.

\begin{figure}
\centerline {\includegraphics[width=0.9\textwidth]{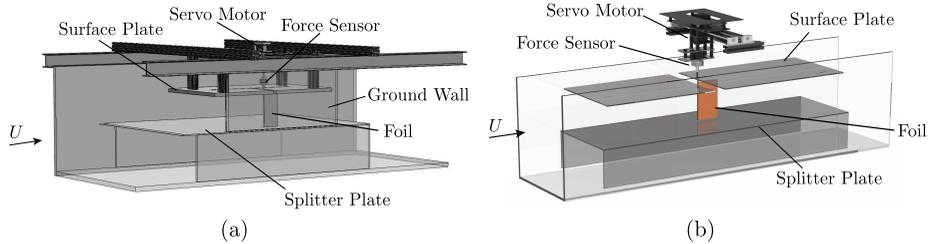}}
\caption{Schematic of (a) the constrained pitching hydrofoil apparatus at Lehigh University, and  (b) the constrained hydrofoil apparatus at the University of Virginia.}
\label{fig:exp-setup}
\end{figure}

\section{Methods}
Potential flow simulations and water-channel experiments were conducted on hydrofoils in and out of ground effect.  The details of the hydrofoil geometry and kinematics, as well as the numerical and experiment methods employed are given below.

\subsection{Hydrofoil Geometry and Kinematics}
The hydrofoil used throughout this study has a rectangular planform shape, a 10\% thick tear-drop cross-section \citep{quinn2014unsteady} with a chord length of $c=0.095$ m, and an effectively infinite aspect ratio. The hydrofoil was actuated with a sinusoidal purely pitching motion about its leading edge of ${\theta}(t)={\theta}_0\sin(2\pi f t)$, where $\theta_0$ is the pitching amplitude, $f$ is the frequency, and $t$ is the time. The frequency defines the reduced frequency, $k \equiv f c/U$, and the Strouhal number, $St \equiv f A/U$.  Here, $A$ is the peak-to-peak amplitude of motion, that is, $A = 2c \sin \theta_0$.  The amplitude of motion is reported in its dimensionless form as $A^* = A/c$.  One of the primary variables of the current study is the dimensionless ground distance, $D^* = D/c$, where $D$ is the distance from the leading-edge of the foil to the ground plane. The input variables used are summarized in Table \ref{TAB:parameters}.

\begin{table}
 \begin{center}
  \begin{tabular}{lccc}
    Variables/Parameters  & Simulations              &   Exp. at UVA (EXP1)        & Exp. at Lehigh U. (EXP2) \\[4pt]
      $A^*=\frac{A}{c}$   &  $0.15 \le A^* \le 0.6 $ &  $0.24 \le A^* \le 0.52 $   & $0.24 \le A^* \le 0.52 $ \\
      $k=\frac{fc}{U}$    & $ 0.1\le k \le 2.0$      & $0.55\le k \le 1.28$      & $0.77\le k \le 1.02$        \\
      $St=\frac{fA}{U}$   & $ 0.15\le St \le 0.60$   & $0.18\le St \le 0.44$        & $0.26\le St \le 0.63$       \\
      $D^*=\frac{D}{c}$   & $ 0.3\le D^* \le 2.0$    & $0.24\le D^* \le 1.665$     & $0.25\le D^* \le 2.6$       \\
      $\AR$               &  $\infty$                & $3$ with end-plates         & $2$ with end-plates         \\
      $\Rey=\frac{U c}{\nu}$  & $\infty$             & $13,$600                    & $9,$000   \\
  \end{tabular}
  \caption{Numerical and experimental variables and parameters.}
  \label{TAB:parameters}
 \end{center}
\end{table}
For the experiments and the simulations, the time-averaged thrust and power coefficients can be non-dimensionalized by the added mass forces and added mass power from small amplitude theory \citep{garrick1936propulsion} or by dynamic pressure,
\begin{small}
\begin{equation}
\label{EQ:CTCP}
    C_T  \equiv \frac{\overline{T}}{\rho S_p f^2 A^2}, \quad C_P  \equiv \frac{\overline{P}}{\rho S_p f^2 A^2 \overline{U}},\quad    C^{\text{dyn}}_T \equiv \frac{\overline{T}}{1/2 \rho S_p U^2}, \quad C^{\text{dyn}}_P \equiv \frac{\overline{P}}{1/2 \rho S_p U^3}.
\end{equation}
\end{small}
\noindent where $\rho$ is the density of the fluid medium and the two normalizations are related through the Strouhal number by simple transformations: $C_T^{\text{dyn}} = C_T\, (2 St^2)$ and $C_P^{\text{dyn}} = C_P \, (2 St^2)$. The propulsive efficiency can be defined as: $\eta\equiv C_T/C_P\equiv C_T^{dyn}/C_P^{dyn}$. 

\subsection{Numerical Method}
To model the potential flow around a foil, we used a two-dimensional boundary element method (BEM) where the flow is assumed to be irrotational, incompressible and inviscid.  By following previous studies \citep{katz2001low,quinn2014unsteady,Moored2018Bem}, the general solution to the potential flow problem is reduced to finding a distribution of sources and doublets on the foil surface and in its wake.  At each time step a no flux boundary condition is enforced on the body.   To solve this problem numerically, constant strength source and doublet line elements are distributed over the body and the wake. Each body element is assigned a collocation point, which is located at the center of each element and shifted 1\% of local thickness into the body where a constant-potential condition is applied to enforce no flux through the surface (i.e. Dirichlet formulation). This results in a matrix representation of the boundary condition that can be solved for the body doublet strengths once a wake shedding model is applied. Additionally, at each time step, a wake boundary element is shed with a strength that is set by applying an explicit Kutta condition, where the vorticity at the trailing edge is set to zero. The presence of the ground is modeled using the method of images, which automatically satisfies the no-flux boundary condition on the ground plane.

A wake roll-up algorithm is implemented at each time step where the wake elements are advected with the local velocity.  During wake roll-up, the point vortices, representing the ends of the wake doublet elements, must be desingularized for numerical stability of the solution \citep{Krasny1986}.  At a cutoff radius of $\epsilon/c=5\times10^{-2}$, the irrotational induced velocities from the point vortices are replaced with a rotational Rankine core model. 

The tangential perturbation velocity component is calculated by local differentiation of the perturbation potential.  The pressure acting on the body is found via applying the unsteady Bernoulli equation. Moreover, the mean thrust force is calculated as the time-average of the streamwise directed pressure forces and the time-averaged power input to the fluid is calculated as the time average of the negative inner product of the force vector and velocity vector of each boundary element, that is, $P = -\int_\mathcal{S} \mathbf{F}_{\text{ele}} \cdot \mathbf{u}_{\text{ele}} \, d\mathcal{S}$ where $\mathcal{S}$ is the body surface.

Convergence studies found that the thrust and efficiency changes by less than 2\% when the number of body panels, $N_{b}= 150$, and the number of time steps per cycle, $N_t= 150$, were doubled independently. The current study considered the foil's cycle-averaged thrust and efficiency as convergence metrics since these are the prime output variables of interest.  The computations were run over 10 flapping cycles and the time-averaged data are obtained by averaging the last cycle. For all simulations there was less then 1\% change in the thrust and efficiency after 7 flapping cycles.  The two-dimensional formulation in the current study has been validated extensively against continuous swimming theory, numerics and experiments \cite[]{quinn2014unsteady,Moored2018Bem,akoz2018unsteady,kurt2019swimming}. For more details on the numerical method see \cite{Moored2018Bem} and \cite{moored2018inviscid}.

\subsection{Experimental Methods}
New experiments (EXP1) were conducted in a closed-loop water channel (Rolling Hills 1520) at the University of Virginia with a foil of aspect ratio $\AR=3$ and a Reynolds number of $\Rey=13,$$600$. A nominally two-dimensional flow was achieved by installing a horizontal splitter plate and a surface plate near the tips of the hydrofoil (Figure \ref{fig:exp-setup}b). The gap between the hydrofoil tips and the surface/splitter plate was less than 5 mm. Surface waves were also minimized by the presence of the surface plate. The wall of the channel was used as a ground plane and the dimensionless ground distance was varied within a range of $0.35 \leq D^* \leq 1.665$. Five different dimensionless amplitudes were tested: $A^* = 0.24$, $0.31$, $0.38$, $0.45$ and $0.52$ resulting in $0.2 \leq St\leq 0.45$.

A second (EXP2) previously published experimental data set \cite[]{kurt2019swimming} is also used for further validation of the proposed scaling laws.  For further experimental details see \cite{kurt2019swimming}.  
The parameters and ranges of variables for both EXP1 and EXP2 can be found in Table \ref{TAB:parameters}.

\section{Scaling laws for a pitching foil in ground effect}
\begin{figure}
\centerline {\includegraphics[width=0.9\textwidth]{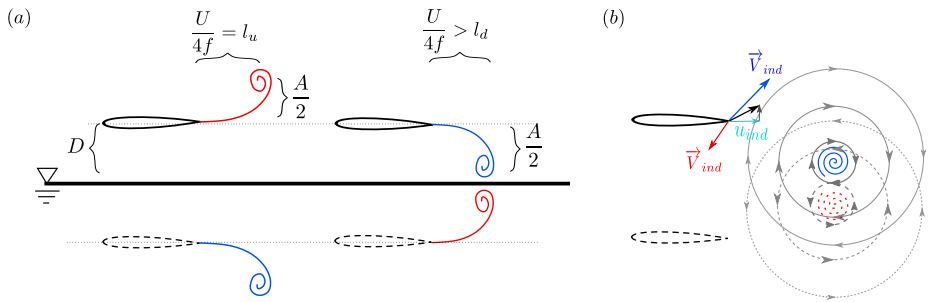}}
\caption{Effects of the image vortex system. (a) The down-stroke negatively-signed vortex slows downs due to the influence of its image vortex ($l_u>l_d$). (b) The induced streamwise velocity of the shedding vortex and its image. The induced velocity from the pair is smaller than a comparable isolated vortex.}
\label{fig:schematic}
\end{figure}

\cite{moored2018inviscid} introduced scaling relations for the performance of two-dimensional, self-propelled pitching hydrofoils. Here, we will briefly review these scaling laws since they will serve as the basis with which to apply novel modifications to the added mass and circulatory forces to account for the proximity to the ground.

\subsection{Scaling laws of an isolated swimmer}
\cite{moored2018inviscid} developed thrust and power scaling laws as a combination of the added mass and circulatory forces from classical linear theory \cite[]{garrick1936propulsion} with additional nonlinear terms that are not accounted for in linear theory. For instance, the thrust coefficient defined in eq. \eqref{EQ:CTCP}, is proposed to be proportional to the superposition of three terms,
\begin{small}
\begin{align}
\label{eq:thrustmoored}
    C_T &= c'_1+ c'_2\phi_2 + c'_3\phi_3 \\ \nonumber
    \text{with:}\quad \phi_2 &= -\biggl[\frac{3F}{2}+\frac{F}{\pi^2 k^2}-\frac{G}{2\pi k}-( F^2+G^2) \biggl(\frac{1}{\pi^2 k^2} + \frac{9}{4} \biggr)  \biggr],\quad \phi_3=A^*,
\end{align}
\end{small}
\noindent where $c'_1$, $c'_2$, and $c'_3$ are constants, and $F$ and $G$ are the real and imaginary components of Theodorsen’s lift deficiency function, respectively \cite[]{theodorsen1935general}. The first and second terms represented by $c'_1$ and $c'_2 \phi_2$ are the added mass and circulatory thrust forces, respectively, from linear theory while the third term represented by $c'_3 \phi_3$ is not accounted for in linear theory. The third term corresponds to the form drag that is proportional to the time varying projected frontal area that occurs during large-amplitude pitching oscillations. \cite{moored2018inviscid} also proposed that the power coefficient defined in equation \eqref{EQ:CTCP} is a linear superposition of three terms as, 
\begin{small}
\begin{align}
\label{eq:powermoored}
    C_P &= c'_4+ c'_5\phi_5 + c'_6\phi_6  \quad\text{with:}\quad \phi_5 = \frac{St^2}{k}\biggl(\frac{k^*}{1+k^*}\biggr),\quad \phi_6=St^2k^*, 
\end{align}
\end{small}
where $c'_4$, $c'_5$, and $c'_6$ are arbitrary constants, and $k^*=k/(1+4St^2)$. The first term ($c'_4$) is the added mass power from linear theory. The second term $(c'_5 \phi_5)$ is a power term that is not present in linear theory and develops from the x-component of velocity of a pitching propulsor, which is neglected in linear theory due to a small-amplitude assumption. For large amplitude motions this velocity does not disappear, leading to an additional velocity component on the bound vorticity of the propulsor and creating an additional contribution to the generalized Kutta-Joukowski force also known as the vortex force (Saffman 1992). The third term $(c'_6 \phi_6)$ is also a power term that is absent in linear theory and develops during large-amplitude motions when the trailing-edge vortices are no longer planar as assumed in the theory. As a result, the proximity of the trailing-edge vortices induce a streamwise velocity over the foil and an additional contribution to the vortex force. In short, the second and third terms are described as the large-amplitude separating shear layer and vortex proximity power terms, respectively, and both terms are circulatory in nature. For more details on the development of the two-dimensional scaling relations see \cite[]{moored2018inviscid}.

\subsection{Scaling laws modifications for ground effect}
In presence of the ground, it is postulated that the added mass (and thus the added mass thrust and power terms), as well as the vortex proximity power term will be affected and will act as the primary drivers for the observed scaling trends with ground proximity. These two modifications for a pitching foil are described below.

The close proximity of a solid boundary can cause a substantial increase in the added mass of a pitching foil. This is due to the increase in fluid acceleration between the foil and boundary \cite[]{brennen1982review}. Classic hydrodynamic theory shows that for a circular cylinder with the radius $r$ moving with a distance $d$ from a solid boundary (when ${d}/{r}<1$) the added mass increases as the cylinder moves closer to the boundary. The added mass can be represented as the addition of the isolated added mass with an additional added mass due to the acceleration of flow in the presence of a solid boundary as,
\begin{small}
\begin{equation}
\label{EQ:addedmassCylinder}
 M_a=\pi\rho r^2 \left[1 +\sum_{j=1}^{\infty} \frac{1}{2^{2j-1}}\left(\frac{r}{d}\right)^{2j}\right] 
\end{equation}
\end{small}
Using conformal mapping, one can directly define the added mass for a specific foil profile \cite[]{korotkin2009added}, however, in this study, the cylinder is mapped to a thin airfoil with a chord length of $c=2r$ for generality. Furthermore, since the dimensionless distance of the foil to the ground is limited by the maximum trailing edge amplitude, we neglect the effect of higher order terms of the series and assume the additional added mass contribution scales as the inverse of the dimensionless distance squared and its coefficient is to be determined. For extreme ground effect problems, the higher order terms in the series are required. The added mass terms of equations \eqref{eq:thrustmoored} and \eqref{eq:powermoored} can then be modified to 
\begin{small}
\begin{align}
\label{EQ:changeaddedmass}
      c'_1=c_1+c_2\zeta_2, \quad c'_4=c_5+c_6\zeta_6 
\end{align}
\end{small}
\noindent where the $\zeta_2=\zeta_6=(1/D^*)^2$ are the additional added mass contributions to the thrust and power generation.

The second mechanism that significantly affects the power generation of a pitching foil in proximity to the ground is circulatory in nature and is best explained using the method of images, where each shedding vortex has an oppositely-signed image vortex in the ground (see Figure \ref{fig:schematic}a).The image vortex of the near-ground negatively-signed vortex induces an additional velocity at the trailing-edge of the foil (see figure~\ref{fig:schematic}b) that acts to reduce the effect of the near-ground vortex and needs to be accounted for in the vortex proximity term in equation \eqref{eq:powermoored}. The power contribution due to the proximity of the vortex is proportional to the additional lift generated from the streamwise induced velocity as $L_\mathrm{Prox}\approx \rho s u_{ind} \Gamma_{w}$ based on Kutta-Joukowski theorem. The image vortex modifies this induced streamwise velocity at the trailing edge as,
\begin{small}
\begin{align}
\label{EQ:Vortexproxcorrection}
u_\mathrm{ind}=\Gamma_w\frac{f}{U}\biggl[\overbrace{\frac{St}{1+4St^2}}^{\text{Near-Ground Vortex}}-\overbrace{2\frac{St+4kD^*}{1+4(4kD^*+St)^2}}^{\text{Image Vortex}}\biggr],
\end{align}
\end{small}
\noindent where the advection speed of the vortex pair is $U$ and the $\Gamma_w$ is the stragnth of the trailing edge vortex circulation. This induced streamwise velocity decreases with decreasing ground distance due to the near nullification of the induced velocity from the near-ground vortex by its image.

The thrust circulatory term is also affected by this image vortex, however, the forces acting on purely pitching foils are dominated by added mass forces, and the circulatory modifications are not significant. The scaling laws for the thrust and power coefficient of a purely pitching foil in proximity to the ground are then,
\begin{small}
\begin{align}
\label{EQ:GFscale}
    C_T &= c_1+ c_2\zeta_2+c_3\zeta_3+c_4\zeta_4, \quad C_P = c_5+ c_6\zeta_6+c_7\zeta_7+c_8\zeta_8 \\ \nonumber
    \text{with:} \quad \zeta_2&=\zeta_6=\frac{1}{{D^*}^2}, \quad\zeta_3=-\biggl[\frac{3F}{2}+\frac{F}{\pi^2 k^2}-\frac{G}{2\pi k}-( F^2+G^2) \biggl(\frac{1}{\pi^2 k^2} + \frac{9}{4} \biggr)  \biggr],\\ \nonumber
    \zeta_4&=A^*, \quad \zeta_7=\frac{St^2}{k}\biggl(\frac{k^*}{1+k^*}\biggr), \quad \zeta_8 = St^2k^*-2k St\frac{(St-4kD^*)}{1+4(St-4kD^*)^2}. 
\end{align}
\end{small}

The scaling laws can also be written in terms of the thrust and power coefficients normalized by dynamic pressure as,
\begin{small}
\begin{align}
\label{EQ:GFScaledyno}
    C_T^\mathrm{dyn} &= 2St^2(c_1+ c_2\zeta_2+c_3\zeta_3+c_4\zeta_4), \quad C_P^\mathrm{dyn} = 2St^2(c_5+ c_6\zeta_6+c_7\zeta_7+c_8\zeta_8).
\end{align}
\end{small}

\begin{figure}
\centerline {\includegraphics[width=0.9\textwidth]{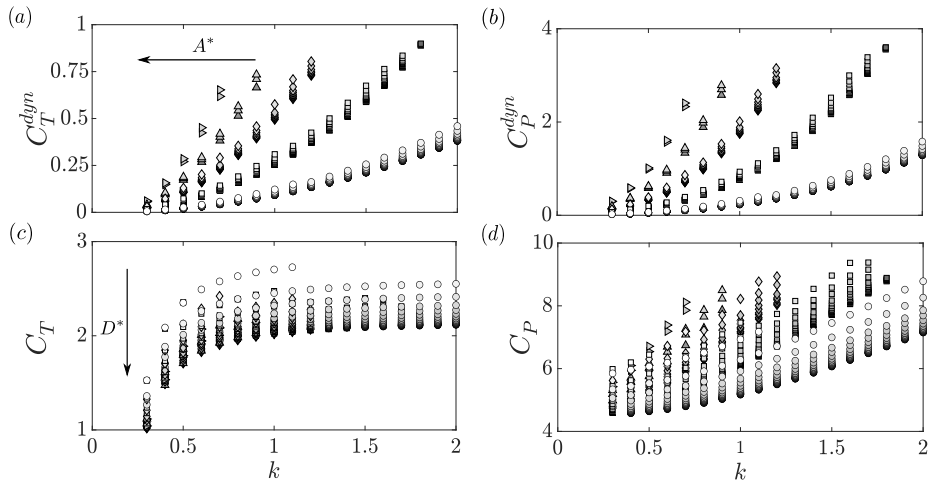}}
\caption{Coefficient of thrust and power as a function of reduced frequency from the self-propelled simulations. The marker colors going from black to white indicate the $D^*$ from far to close the ground, respectively, over the range $0.3 \leq D^* \leq 2.0$. (a) and (c) normalized based on dynamic pressure (b) and (d) normalized based on added mass force. }
\label{fig:coeff_num}
\end{figure}

\section{Results and Discussion}
The combination of computational input variables presented in Table \ref{TAB:parameters} leads to 665 two-dimensional simulations with a Strouhal number range of $0.1  \leq St  \leq 0.6$ and a reduced frequency range of $0.1 \leq k \leq 2.0$.  From these simulations, the thrust and power coefficients as defined in equation \eqref{EQ:CTCP}  are presented in Figure \ref{fig:coeff_num}. As in previous work \cite[]{quinn2014unsteady}, thrust and power coefficients increase with amplitude and ground proximity (Figure \ref{fig:coeff_num}).

\begin{figure}
\centerline {\includegraphics[width=0.9\textwidth]{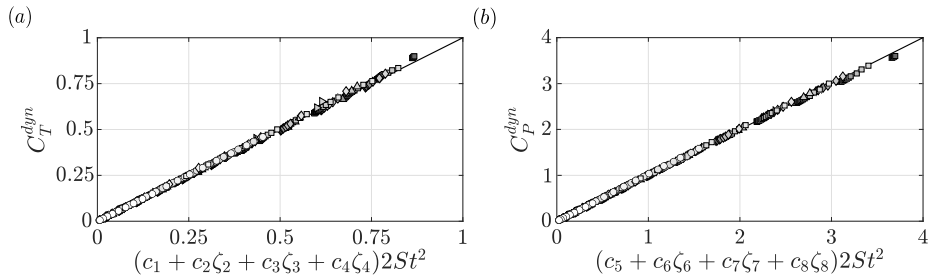}}
\caption{Scaling of the (a) thrust and (b) power coefficients for all motion amplitudes and ground proximities considered in the numerical simulations.}
\label{fig:scale_num}
\end{figure}

Figure  \ref{fig:scale_num}  presents  the  numerical  data  plotted  as  a  function  of  the ground effect scaling laws proposed in eqs. (\ref{EQ:GFscale}) and (\ref{EQ:GFScaledyno}).  An excellent collapse of the data is observed showing that the scaling laws capture the physics of unsteady ground effect in potential flows.  The collapsed data can be seen to follow a line of slope one for both the thrust and power within $\pm 2\%$ of the predicted scaling law.  The  constants in the thrust law are $c_1 = 2.99$, $c_2 = 0.06$, $c_3 =-4.43$ and $c_4 = -0.09$, while for the power law they are $c_5 = 4.46$, $c_6 = 0.14$, $c_7 = 25.2$ and $c_8 = 14.13$.

\begin{figure}
\centerline {\includegraphics[width=0.9\textwidth]{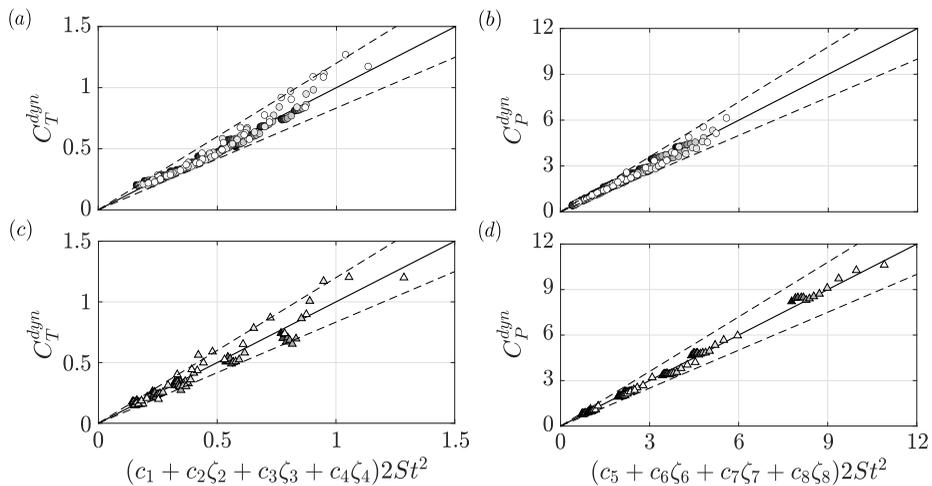}}
\caption{Scaling of the time averaged thrust and power for (a-b) UVA data set(EXP1)  (c-d)LU data set(EXP2) for all motion amplitudes and distances from the wall considered in experiments. The dashed lines presents 20\% margins of error. }
\label{fig:scale_exp}
\end{figure}

To validate whether the scaling laws can also apply to viscous flows, two experimental data sets are graphed against the scaling law predictions in Figure \ref{fig:scale_exp}. The first experimental data set (EXP1) shows a collapse of the data to within $\pm 20 \%$ (thrust and power) of the scaling law prediction.  The deviation of both thrust and power from the scaling law prediction is likely a result of bending of the actuation rods at high $St$, which introduces a heaving component to the foil motion that is not accounted for in the scaling laws. The experimentally determined coefficients for the thrust are $c_1 = 5.03$, $c_2 = 0.08$, $c_3 = -5.72$, $c_4 = -4.66$ and for the power they are $c_5 = 4.313$, $c_6 =0.004$, $c_7 = 0.11$ and $c_{8} = 30.91$. The second experimental data set (EXP2) shows a collapse within $\pm 22.5 \%$ (thrust) and $\pm 11.5 \%$ (power) of the scaling law prediction.  The higher error margin in the thrust scaling is attributed to secondary viscous effects not accounted for in the scaling laws. The experimentally determined coefficients for the thrust law are $c_1 = 4.29$, $c_2 = 0.09$, $c_3 = -14.33$, $c_4 = -0.52$, and for the power law they are $c_5 = 8.26$, $c_6 = 0.03$, $c_7 = 33.06$ and $c_{8} = 15.38$.

Our scaling relations show a good collapse of the data for a wide range of Reynolds number from $\Rey=9,$$000$ and $\Rey = 13,$$600$ in the experiments to $Re = \infty$ in the inviscid simulations. Although, it should be noted that the determined coefficients are different among the experimental and numerical data sets, which highlights that the coefficients likely vary with $Re$ as observed by \cite{Senturk2019}.  In support of the proposed scaling laws the added mass thrust and power terms are positive in both experiments and simulations, as expected based on physical grounds. Although it is clear that the Reynolds number can alter the coefficients, no additional terms need to be introduced to account for data obtained at different $Re$.  This supports the previous conclusion \cite[]{kurt2019swimming} that the dominant flow physics in ground effect are inviscid in nature. The small differences between the scaling law agreement in the experiments and the simulations may be attributed to secondary viscous effects. 

The collapse of the data to a line of slope one for both numerical and experimental cases confirms that the newly proposed scaling laws capture the dominant flow physics of two-dimensional pitching propulsors in ground effect across a wide range of $St$, $A^*$ and $D^*$.

\section{Conclusion}
New scaling laws are developed for the thrust generation and power consumption of two-dimensional pitching propulsors in ground-effect by extending the two-dimensional pitching scaling laws introduced by \cite{moored2018inviscid} to consider added mass and circulatory effects due to the close proximity of a ground plane. The developed scaling laws are shown to predict inviscid numerical data and experimental data well, within $\pm$20\% of the thrust and power data, respectively. The scaling laws reveal that both an increase in the added mass with decreasing ground distance and a reduction in the influence of a shed vortex by its image with decreasing ground distance are key physics to capture in a scaling law that is valid over a wide range of ground distances, motion amplitudes, and Strouhal numbers.  These results can be extended to two swimmers in side-by-side arrangements with an out-of-phase synchronization. The established scaling relationships elucidate the dominant flow physics behind the force production and energetics of pitching bio-propulsors and can be used to accelerate the design of
bio-inspired devices that swim near a ground plane and operate in side-by-side schools.  

\section{Acknowledgements}
This work was supported by the Office of Naval Research under Program Director Dr. Robert Brizzolara on MURI grant number N00014-08-1-0642 and BAA grant number N00014-18-1-2537, as well as by the National Science Foundation under Program Director Dr. Ronald Joslin in Fluid Dynamics within CBET on NSF CAREER award number 1653181 and NSF collaboration award number 1921809.

\bibliographystyle{jfm}
\bibliography{jfm-groundeffscale}

\end{document}